# A data driven approach for cross-slip modelling in continuum dislocation dynamics

Vignesh Vivekanandan[1], Ben Anglin[2], Anter El-Azab[1]

[1] School of Materials Engineering, Purdue University, Neil Armstrong Hall of Engineering, 701 W Stadium Avenue, West Lafayette, IN 47907, USA

[2] Naval Nuclear Laboratory, 814 Pittsburgh-McKeesport Blvd, West Mifflin, PA 15122, USA

**Abstract**

Cross-slip is a thermally activated process by which screw dislocation changes its glide plane to another slip plane sharing the same Burgers vector. The rate at which this process happens is determined by a Boltzmann type expression that is a function of the screw segment length and the stress acting on the dislocation. In continuum dislocation dynamics (CDD), the information regarding the length of the screw dislocation segment and local stress state on dislocations are lost due to the coarse-grained representation of the density. In this work, a data driven approach to characterize the lost information by analyzing the discrete dislocation configurations is proposed to enable cross-slip modeling in the CDD framework in terms of the coarse-grained dislocation density and stress fields. The analysis showed that the screw segment length follows an exponential distribution, and the stress fluctuations, defined as the difference between the stress on the dislocations and the mean field stress in CDD, follows a Lorentzian distribution. A novel approach for cross slip implementation in CDD employing the screw segment length and stress fluctuation statistics was proposed and rigorously tested by comparing the CDD cross-slip rates with discrete dislocation dynamics (DDD) rates. This approach has been applied in conjunction with three cross-slip models used in DDD simulations differing mainly in the functional form of cross slip activation energy. It was found that different cross-slip activation energy formulations yielded different cross-slip rates, yet the effect on mechanical stress-strain response and dislocation density evolution was minimal for the [001] type loading.

# 1. Introduction

Predicting the self-organized dislocation microstructure is considered a holy grail of the dislocation theory, which is currently belied to be potentially achievable through continuum dislocation dynamics (CDD) models (Hochrainer, 2015; Kalaei et al., 2022; Leung and Ngan, 2016; Lin et al., 2021; Lin and El-Azab, 2020; Monavari and Zaiser, 2018; Sandfeld et al., 2015; Schulz et al., 2019; Starkey and El-Azab, 2022; Sudmanns et al., 2019; Vivekanandan et al., 2021; Xia and El-Azab, 2015) of mesoscale plasticity. Achieving this goal requires modelling the dislocation mechanisms within the CDD framework. One such important mechanism is cross-slip, which is a thermally activated process in which the screw dislocations change their glide plane. It is well established that cross slip contributes to pattern formation (Kubin et al., 2009), dynamic recovery (Madec et al., 2002) and strain hardening (Devincre et al., 2007) in metals. Therefore, it is essential for continuum dislocation dynamics models to accurately incorporate cross-slip mechanism to aid the prediction of dislocation microstructure evolution.

Broadly, there exists two types of models that explain the mechanism of cross-slip, Friedel-Escaig mechanism (FE) (Bonneville and Escaig, 1979) and Fleischer mechanism (FL) (Fleischer, 1959). The FE mechanism postulates that cross-slip happens in, say, FCC crystals when the two dislocation partials form a constriction on the glide plane and dissociate onto the cross-slip plane whereas the FL mechanism predicts the leading partial splits into a partial dislocation on the cross-slip plane and a stair-rod dislocation which eventually reacts with the trailing partial and completes the cross-slip process. Capturing the rate at which this process happens depends upon the activation energy barrier of the critical configuration and the length of the screw segment. Hence, it is essential for continuum dislocation dynamics models to incorporate these two parameters for accurately predicting the cross-slip behavior.

Continuum dislocation dynamics models represent dislocations as coarse-grained density fields which have information about the average line orientation and magnitude of the dislocation density. The advantage of using this spatially smooth variable in studying the dynamics is the evolution of the system can be encapsulated in terms of the average quantity, e.g., dislocation density, rather than tracking the motion of all discrete states (Bertin et al., 2015; Cui et al., 2015; Devincre et al., 2011; Hussein et al., 2015; Po et al., 2014; Sills et al., 2018; Stricker et al., 2018; Zhou and LeSar, 2012) that make the density. Conversely, coarse graining throws away the details corresponding to the discrete states like dislocation segment length distribution and dislocation-dislocation correlation which are important parameters required to model local processes like cross-slip. Hence, for the coarse-grained models to remain consistent with their lower-level counterparts, care must be taken to introduce the statistical information regarding the cross-slip process.

There exist different classes of continuum dislocation based models in the literature (Acharya and Roy, 2006; Hochrainer, 2015; Kalaei et al., 2022; Li et al., 2014; Lin et al., 2021; Lin and El-Azab, 2020; Monavari and Zaiser, 2018; Sandfeld et al., 2015; Schulz et al., 2019; Starkey and El-Azab, 2022; Sudmanns et al., 2019; Vivekanandan et al., 2021; Xia and El-Azab, 2015). Here, we focus upon the vector-based CDD models (Lin et al., 2021; Lin and El-Azab, 2020; Starkey and El-Azab, 2022; Vivekanandan et al., 2021; Xia and El-Azab, 2015) and higher order continuum models (Hochrainer, 2015; Sandfeld et al., 2015; Sudmanns et al., 2019) that explicitly capture the cross-slip process. The first attempt to characterize the density-based rates of cross-slip process by systematically studying it using DDD (Devincre et al., 2011) was done by (Deng and El-Azab, 2010), which was then improved by (Xia et al., 2016). In the latter work, attempts were made to characterize the cross-slip process for vector based CDD models by temporally coarse graining the time series data of cross-slip statistics obtained during every time step of the DDD simulations. The cross-slip statistics used in the time series were obtained by monitoring the fraction of screw segments that cross-slipped throughout the DDD simulation domain in each time step. These coarse-grained time series were then used to sample cross-slip rates in CDD based on the strain levels. Another approach to modeling cross-slip was proposed by (Sudmanns et al., 2019) in which they used the DDD cross-slip probability expression (Kubin et al., 1992) to calculate the cross-slip rate. In their work, the length of screw segment allowed to cross-slip, which is an input to the cross-slip probability expression, was estimated as the minimum of the average loop radius, defined as the fraction of dislocation density and curvature density, and the mean dislocation spacing. Although both the models make use of information from DDD, there is still a scope for improvement. For instance, the model proposed by (Deng and El-Azab, 2010) and (Xia et al., 2016) does not consider the local stress effects to estimate the CDD cross-slip rate but rather uses the average cross-slip rate information obtained from DDD. Similarly, the model proposed by Sudmanns.et.al (Sudmanns et al., 2019) does not consider the distribution of screw segment length rather estimates the average length based on the coarse grained fields for cross-slip probability estimation. In addition to that, it was recently pointed out that the stress faced by a discrete dislocation is markedly different from the mean stress field (Vivekanandan et al., 2022), which is not accounted for in their cross-slip probability evaluation. Hence, there is a need for a different approach which relates the coarse-grained dislocation density and stress fields in CDD to the discrete screw segment length distribution and stress field on dislocations to obtain the cross-slip rates accurately.

Since cross-slip is a thermally activated process, the estimation of the activation energy barrier required for the screw segment to transition from glide plane to cross-slip plane is the crucial step. Predominantly, two different approaches have been considered in the literature to estimate the activation energy barrier; the line tension models (Kang et al., 2014; Malka-Markovitz and Mordehai, 2018) and atomistic simulation models (Kuykendall et al., 2020; Rao et al., 2013). Based on these two approaches, different activation energy expressions were derived and used in DDD simulations. Most commonly used activation energy expression was the one first proposed by (Kubin et al., 1992) and later modified by (Devincre et al., 2011) which states that activation energy of cross-slip barrier mainly depends on the Schmid stress along the primary plane ($\tau_s^p$) for an immobile dislocation and Schmid stress along cross-slip plane ($\tau_s^{cs}$) for a mobile dislocation. The parameters used in this model were calibrated based on the experimental observations at the initiation of stage III hardening resulting in activation energy of the form $E_{act} = -V(|\tau_s^p| - \tau_{III})$ or $E_{act} = -V(|\tau_s^{cs}| - \tau_{III})$ depending on whether the dislocation is immobile or mobile, where $V$ is the activation volume and $\tau_{III}$ is the resolved shear stress at the beginning of stage III hardening. Contrary to the Kubin's model, some studies (Kang et al., 2014) showed that the effect of Escaig stress on the activation barrier is much more significant than the Schmid stress. Based on atomistic studies conducted by Rao and co-workers (Rao et al., 2013, 2011, 2010), Hussein and co-workers (Hussein et al., 2015) proposed a new cross-slip model in which the activation energy of cross-slip barrier was assumed to be linearly dependent on the difference in Escaig stress in the primary plane $(\tau_e^p)$ and cross-slip plane $(\tau_e^{cs})$ as follows: $E_{act} = E_{cons} - V(\tau_e^p - \tau_e^{cs})$ where $E_{cons}$ is the constriction energy. Apart from these two models, Malka-Markovitz and Mordehai (Malka-Markovitz and Mordehai, 2019, 2018) derived a closed form expression for the stress dependent activation energy barrier based on the line tension model by employing harmonic approximation to obtain the interaction energy between partial dislocations. In this model, the contributions of both Schmid stress and Escaig stress in glide and cross-slip plane were considered in the activation energy expression.

In this work, a new cross-slip model for vector-based CDD model is proposed. This model incorporates information about the screw segment statistics and internal stress fluctuations. Section 2 introduces the fundamental equations of the CDD model which is followed in section 3 by a discussion of the strategy for estimating the cross-slip rate using DDD statistics in terms of coarse-grained fields. In section 4, the validity of the proposed model is demonstrated by comparing the results with the cross-slip rates evaluated using the DDD. The effects of the different cross-slip activation energy formulations on the mechanical behavior and dislocation microstructure evolution in CDD is also discussed in this section. We conclude by discussing the results and summarizing the conclusions in Section 5 and 6.

## 2. Continuum dislocation dynamics model

Continuum dislocation dynamics models generally include two sets of equations describing the crystal mechanics and dislocation kinetics, which, together, capture the spatial-temporal evolution of dislocation system and the deformation and mechanical state of the crystal under mechanical boundary conditions. The crystal mechanics part is modelled as an eigen distortion problem and dislocation kinetics is captured in terms of a curl type dislocation transport reaction coupled with dislocation processes like cross-slip and junction reactions. The governing equations for the crystal mechanics include the stress equilibrium equation, Hooke's law and boundary conditions:

$$\nabla \cdot \boldsymbol{\sigma} = \boldsymbol{0} \text{ in } \Omega, \tag{1}$$

$$\boldsymbol{\sigma} = \boldsymbol{C} : (\boldsymbol{\beta}^e)_{\text{sym}} \text{ in } \Omega,$$

$$\boldsymbol{\beta}^e = \nabla \boldsymbol{u} - \boldsymbol{\beta}^p,$$

$$\boldsymbol{u} = \bar{\boldsymbol{u}} \text{ on } \partial\Omega_u,$$

$$\boldsymbol{n} \cdot \boldsymbol{\sigma} = \bar{\boldsymbol{t}} \text{ on } \partial\Omega_t.$$

In the above equation, $\boldsymbol{\sigma}$ is Cauchy stress tensor, $\boldsymbol{C}$ is elastic tensor, $\boldsymbol{u}$ is the displacement, $\boldsymbol{\beta}^e$ and $\boldsymbol{\beta}^p$ are elastic and plastic distortions, respectively, $\Omega$ is the domain of solution, $\partial\Omega_u$ and $\partial\Omega_t$, are the parts of the boundary on which the displacement $\boldsymbol{u} = \bar{\boldsymbol{u}}$ and the traction $\bar{\boldsymbol{t}}$ are prescribed respectively. Solution of the above boundary value problem yields the stress which is then used to determine the dislocation velocity for each slip system, $i$, using a linearized mobility law commonly used in literature (Lin and El-Azab, 2020; Yefimov et al., 2004; Zaiser et al., 2001) as follows

$$v^i = \text{sgn}(\tau^i) \frac{b}{B} \langle |\tau|^i - (\tau_0 + \tau_p^i) \rangle, \tag{2}$$

where $b$ and $B$ are the magnitude of the Burgers vector and the drag coefficient, respectively. The $\langle \cdot \rangle$ corresponds to the Macauley bracket and sgn function corresponds to the sign of the argument. $\tau^i$ is the resolved shear stress along slip system $i$ and $\tau_0$ is the lattice friction. $\tau_p^i$ is the Taylor hardening stress that accounts for short range interactions due to sessile junctions on slip system $i$ which is defined as

$$\tau_p^i = \alpha\mu b \sqrt{\sum_j a_{ij}\rho^j} \, F(\rho^i, \rho^f). \tag{3}$$

In the above, $\alpha$ is scaling factor, $\mu$ is the shear modulus, $b$ is the magnitude of Burgers vector, $a_{ij}$ is the average strength of an interaction between slip system $i$ and $j$, $\rho^j$ is the scalar dislocation density of the slip system $j$ interacting with slip system $i$ and $\rho^f$ is the sum of the dislocation densities of the slip systems that react with slip system $i$ and form Lomer or Hirth lock. $F(\rho^i, \rho^f)$ corresponds to a function that accounts for the dislocation pile up effect (Vivekanandan et al., 2021; Zhu et al., 2016). In this work, $F(\rho^i, \rho^f)$ is assumed to have the form $F(\rho^i, \rho^f) = \left(1 + pe^{(q\rho^i/\rho^f - 1)}\right)^{-1}$ based on the work of (Vivekanandan et al., 2021), where $p$ and $q$ are parameters that determine the shape and the rate at which the function decays.

The dislocation velocity obtained from Eq. (2) for every slip system $i$ is used to solve the dislocation kinetics equations described below, following the approach used in (Vivekanandan et al., 2021),

$$\dot{\boldsymbol{\rho}}_g^i = \boldsymbol{\nabla} \times (\boldsymbol{v}^i \times \dot{\boldsymbol{\rho}}_g^i) + \dot{\boldsymbol{\rho}}_{cs}^i + \dot{\boldsymbol{\rho}}_{jun}^i, \qquad (4)$$

$$\dot{\boldsymbol{\rho}}_v^i = \dot{\boldsymbol{\rho}}_{v_{cs}}^i + \dot{\boldsymbol{\rho}}_{v_{jun}}^i,$$

$$\boldsymbol{\nabla} \cdot (\boldsymbol{\rho}_g^i + \boldsymbol{\rho}_v^i) = 0.$$

In the above system of equations, the first equation corresponds to the change in glide dislocation density due to dislocation transport and dislocation reactions. The curl type term captures the change in glide dislocation density due to dislocation transport. $\dot{\boldsymbol{\rho}}_{cs}^i$ captures the change in glide dislocation density due to cross-slip which accounts for both new dislocations added to the slip system $i$ as well as dislocations that was removed from slip system $i$. $\dot{\boldsymbol{\rho}}_{jun}^i$ corresponds to the change in glide dislocation density due to formation and destruction of dislocation junction reactions like glissile junction, Lomer locks and Hirth locks. The second equation corresponds to the rate of change of virtual dislocation density, which is defined as the dislocation density which are involved in dislocation reactions like cross-slip and junctions and are no longer available for dislocation glide. It is important to note that virtual density is a place holder used to keep track of dislocations involved in dislocation reactions and does not contribute to the distortion of the crystal. $\dot{\boldsymbol{\rho}}_{v_{cs}}^i$ captures the change in virtual dislocation density due to cross-slip and $\dot{\boldsymbol{\rho}}_{v_{jun}}^i$ corresponds to the change in virtual dislocation density due to formation and destruction of dislocation junction reactions like glissile junction. The third equation corresponds to the dislocation closure constraint that ensures dislocations do not end inside the crystal given that the initial dislocation configuration is divergence free.

Modelling the evolution of dislocations involved in dislocation processes like junctions and cross-slip requires information about the rate at which these processes occur. Since dislocations in CDD are represented as coarse grained fields, they inherently lack the local information about the individual dislocations involved in these processes. In this paper, a new approach based on our recent work (Vivekanandan et al., 2022) is formulated to model cross-slip behavior in CDD.

### 3. Cross-slip rate estimation in CDD

Cross-slip is a thermally activated process by which a screw dislocation changes its original glide plane to another glide plane which shares the same Burgers vector. In CDD, the change in dislocation density due to this process for slip system $i$ is denoted by $\dot{\boldsymbol{\rho}}_{cs}^{i}$ in Eq. (4) which can be defined as

$$\dot{\boldsymbol{\rho}}_{cs}^{i} = \dot{r}_{cs}^{j \to i} \boldsymbol{\rho}_{s}^{j} - \dot{r}_{cs}^{i \to j} \boldsymbol{\rho}_{s}^{i}, \tag{5}$$

where $\boldsymbol{\rho}_{s}^{i}$ and $\boldsymbol{\rho}_{s}^{j}$ denote screw dislocation densities in slip system $i$ and $j$, $\dot{r}_{cs}^{i \to j}$ and $\dot{r}_{cs}^{j \to i}$ correspond to the cross-slip rates of screw segments that cross-slipped from $i \to j$ and $j \to i$ respectively. It is evident from Eq. (5) that the cross-slip rates $\dot{r}_{cs}$ should be determined accurately to characterize the cross-slip process in CDD. In DDD models, the probability for a discrete screw segment of length $L$ to cross-slip is estimated using an Arrhenius type equation (Hussein et al., 2015; Kubin et al., 1992; Malka-Markovitz et al., 2021) as follows

$$P(L, \boldsymbol{\sigma}) = cL e^{-\frac{E_{act}(\boldsymbol{\sigma})}{kT}}, \tag{6}$$

where $c$ is normalization constant, $E_{act}(\boldsymbol{\sigma})$ is the cross-slip activation energy barrier which depends on the local stress state of the dislocation segment, $k$ is the Boltzmann constant and $T$ is the temperature. Based on the cross-slip probability value, the cross-slip event is executed following the Metropolis algorithm. Then, the average cross-slip rate for a given time step over the entire simulation domain for the discrete case can be written as

$$\dot{r}_{\text{cs,DDD}}^{i \to j} = \frac{\sum l_{\text{cs,screw}}^{i}}{\sum l_{\text{screw}}^{i}} \cdot \frac{1}{\Delta t}, \tag{7}$$

where $l_{\text{cs,screw}}$ and $l_{\text{screw}}$ correspond to the length of screw segments belonging to slip system $i$ that cross-slipped to slip system $j$ and length of screw segments belonging to slip system $i$ respectively and $\Delta t$ is the time step.

In CDD, the dislocation content is represented as density field which is a coarse-grained representation of the discrete dislocation. For the case of vector density based CDD model (Lin and El-Azab, 2020; Vivekanandan et al., 2021; Xia and El-Azab, 2015), the resolution over which the coarse graining is performed is small enough such that the underlying discrete dislocations can be assumed to have the same line orientation as the density field. Therefore, if the coarse-grained dislocation density field in CDD is of screw character, then all the underlying discrete dislocations is of screw character as well. It is important to note that this might not be true for other continuum models (Hochrainer, 2015; Sudmanns et al., 2019) where at any given point, dislocations of different orientations can coexist. Since cross-slip is a local process that depends on the discrete length of the screw segment and the stress on the screw segment, a link must be established between the CDD dislocation density field and the underlying discrete dislocation states to estimate the cross-slip rate.

The first step in establishing such a mapping between the coarse-grained CDD dislocation state and the underlying discrete dislocation states is determining the length distribution of discrete dislocation segments in DDD. The length distribution of the screw segments is analyzed by taking snapshots of the DDD simulation at different strain levels. Fig. 1(a) shows the screw segment length distribution of a DDD configuration throughout the simulation domain. It is evident from the figure that screw segment length in DDD follows an exponential distribution whose functional form is given by $S(x; \phi) = \frac{1}{\phi} e^{-\frac{x}{\phi}}$, where $\phi$ is the mean of the distribution. Similar type of distribution has been observed in (Sills et al., 2018), wherein they studied the link length distribution, which refers to length of dislocation between two pinning points in DDD. It is also important to note that the screw segments whose length is below 25nm are ignored in the distribution. The reason behind this exclusion is that in DDD cross-slip for small screw segments is not worth considering since they do not evolve on the cross-slip plane after cross-slipping; rather they cross-slip back again into the original glide plane after some short time. Fig. 1(b) shows the exponential fits of screw segment length distribution across different strain levels. From the curve fits, we can observe that the average screw segment length decreases with increase in strain. This trend is expected because the number of dislocations in the simulation domain increases as the strain increases and consequently chances of collisions between dislocations resulting in junctions also increases. This results in long screw segments splitting into two smaller segments, thereby reducing the mean length of the screw segments in the simulation domain. Furthermore, it was found that Lambda defined as $\Lambda = \frac{1}{\bar{l}}$, where $\bar{l}$ is the mean screw segment length, increases linearly with strain as shown in Fig. 1(c). Based on the statistics of screw segment length, the average screw segment length $\bar{l}$ at every point in CDD can be estimated by sampling value from the exponential distribution for a given strain level. Therefore, for given screw dislocation density $\rho_s$, the number of underlying screw segments per unit volume with an average length $\bar{l}$ can then be defined as $N = \frac{\rho_s}{\bar{l}}$.

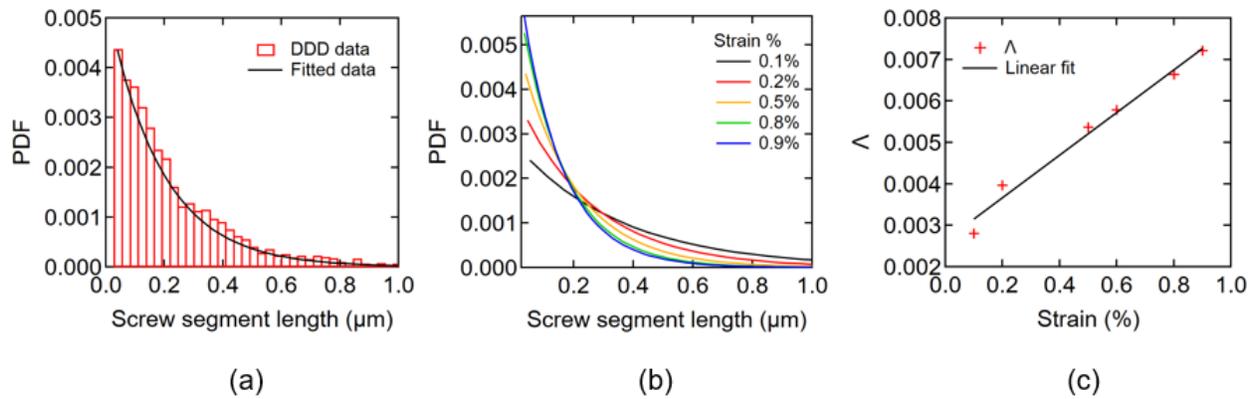

Fig. 1. (a) Length distribution of screw segments for a dislocation configuration sampled from a monotonic DDD simulation at 0.5% strain. (b) Exponential fits of screw segment length distribution at different strain levels. (c) Mean of the exponential fits in (b) fitted to a line.

The second step in establishing a mapping between the coarse-grained CDD dislocation state and the underlying discrete dislocation states is characterizing the local stress state of the underlying screw dislocation in terms of the coarse-grained stress field. Since dislocations are represented as coarse grained density in CDD, the local information regarding the stress state of individual underlying dislocations is lost. Hence, there is a need to formulate a method to estimate the local stress state in terms of coarse-grained stress. The earlier work of a subgroup of authors (Vivekanandan et al., 2022) studied the relationship between the two quantities by analyzing DDD configurations for segments of all orientations. Since cross-slip is relevant only for screw dislocations, the analysis must be performed again exclusively for screw segments. Therefore, a set of discrete dislocation configurations is analyzed to determine the relationship between the local stress and the coarse-grained stress for screw segments. The analysis involves performing coarse graining over voxels for a discrete dislocation configuration, which are regions of small volumes commensurate with the CDD mesh size, thereby converting the discrete states to continuum fields. The details regarding the coarse graining process and stress calculation can be found in our earlier work (Vivekanandan et al., 2022).

Following the steps mentioned in (Vivekanandan et al., 2022), the relationship between the stress on the dislocation line and the coarse grained stress in the voxel is characterized in terms of the scaled stress fluctuation $\widehat{\boldsymbol{\sigma}}^\text{d}$ which is given by the following equation

$$\widehat{\boldsymbol{\sigma}}^\text{d}(\boldsymbol{r}'_k) = \frac{\boldsymbol{\sigma}^\text{d}(\boldsymbol{r}'_k) - [\boldsymbol{\sigma}^\text{c}]_{\lfloor r_k \rfloor}}{[\boldsymbol{\sigma}^\text{c}]_{\lfloor r_k \rfloor}} \quad \text{for } \boldsymbol{r}'_k \in \Lambda \text{ and } \boldsymbol{r}_k \in \text{S}, \tag{8}$$

where $\boldsymbol{\sigma}^\text{d}(\boldsymbol{r}'_k)$ represents the stress on a dislocation line located at the point with position vector $\boldsymbol{r}'_k$ within the voxel, $[\boldsymbol{\sigma}^\text{c}]_{\lfloor r_k \rfloor}$ represents the average stress within a voxel which is estimated by taking the average stress value of all crystal points $\lfloor r_k \rfloor$ located within the voxel. $\Lambda$ and S corresponds to the set of points on the dislocation and set of points on the crystal respectively. For the case of cross-slip, the stress fluctuation statistics of two components of stress tensor $\boldsymbol{\sigma}$, namely Schmid and Escaig are relevant. Schmid stress $\tau_\text{s}$ is the resolved shear stress that drives the dislocation motion while Escaig stress $\tau_\text{e}$ is the stress component that widens or shrinks the stacking fault between partial dislocations. If the Burgers vector of a given slip system $i$ is defined as $b\hat{\boldsymbol{b}}^i$, with $b$ being its magnitude, the slip plane normal be $\hat{\boldsymbol{n}}^i$, and the direction of glide vector be $\hat{\boldsymbol{d}}^i = \hat{\boldsymbol{n}}^i \times \hat{\boldsymbol{b}}^i$, then the Schmid and Escaig components along a slip system $i$ can then be defined as

$$\tau_s^i = \left((\boldsymbol{\sigma} \cdot \hat{\boldsymbol{b}}^i) \times \hat{\boldsymbol{b}}^i\right) \cdot \hat{\boldsymbol{d}}^i, \tag{9}$$

$$\tau_e^i = -\left((\boldsymbol{\sigma} \cdot \hat{\boldsymbol{d}}^i) \times \hat{\boldsymbol{b}}^i\right) \cdot \hat{\boldsymbol{d}}^i. \tag{10}$$

The scaled fluctuation statistics of Schmid and Escaig component of the stress tensor estimated based on Eq. (8) on the screw dislocations throughout the simulation domain is cast in the form of probability distribution function (PDF) as shown in Fig. 2. It is evident from the figure that the PDF follows Lorentzian distribution whose functional form is $H(x: x_0, w) = \frac{1}{\pi}\left(\frac{w}{(x-x_0)^2+w^2}\right)$ where $x_0$ is the center of the distribution and $w$ is the full width at the half maximum. Fig. 3 shows the variation of the scaled fluctuation statistics with respect to the strain level. It can be observed that the peak of the distribution decreases and width of the distribution increases as the strain level is increased. The results of fluctuation statistics for screw segments follows the similar trend with minor differences discussed in (Vivekanandan et al., 2022) which studied dislocations of all orientations.

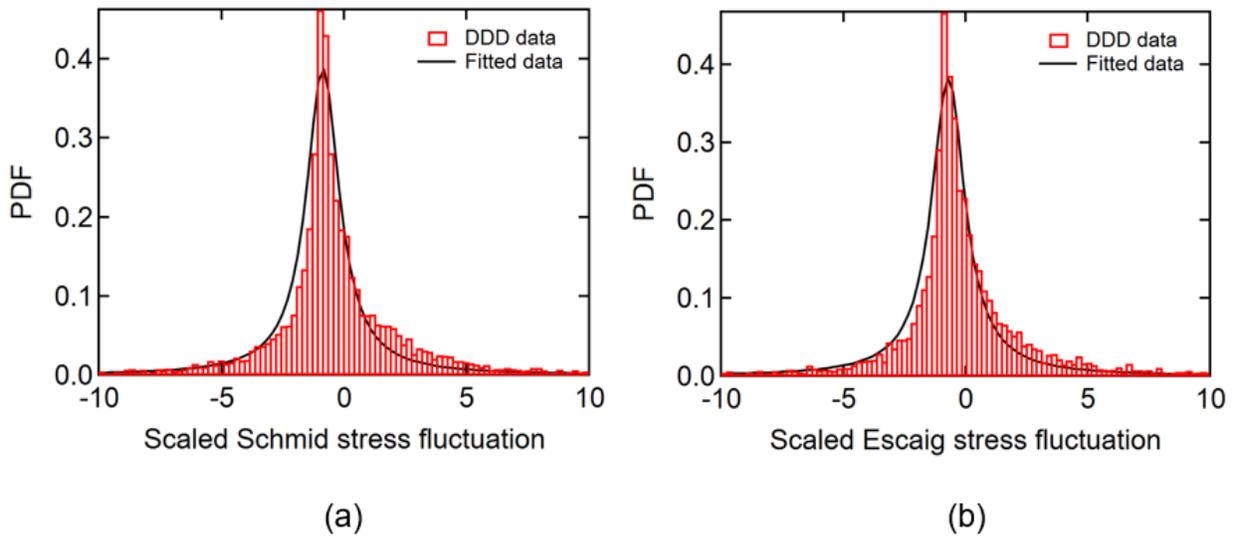

(a)            (b)

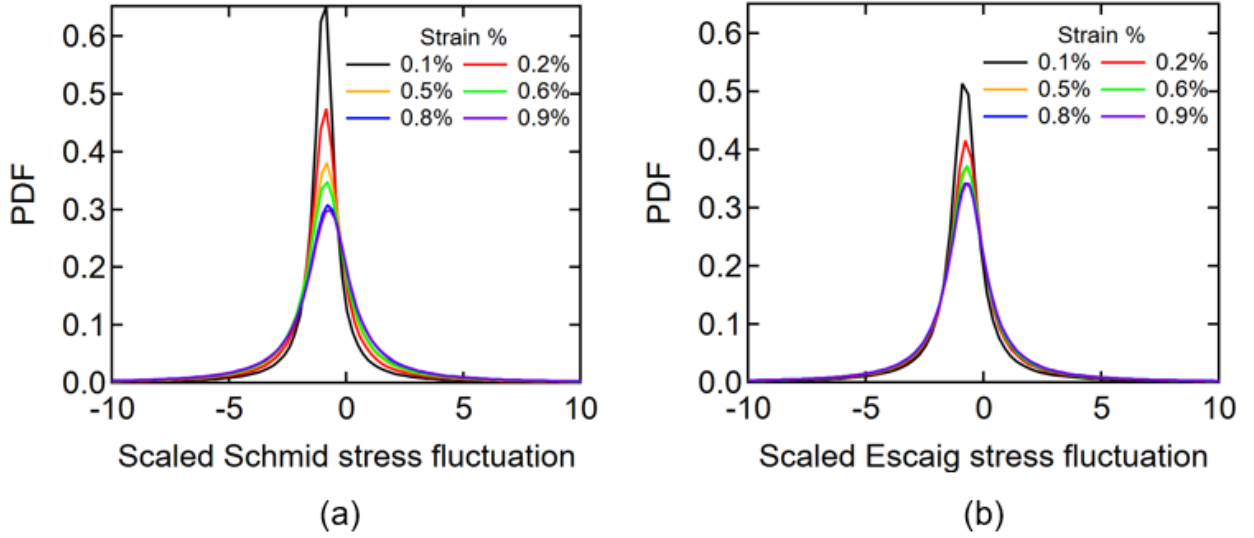

Fig. 2. PDF of (a) scaled Schmid stress fluctuation and (b) scaled Escaig stress fluctuation on dislocation points for screw dislocation segments at 0.5% strain. The statistics of both Schmid and Escaig stress fluctuations follows Cauchy distribution.

Fig. 3. PDF of (a) scaled Schmid stress fluctuation and (b) scaled Escaig stress fluctuation on dislocation points at different strain levels for screw dislocation segments.

Based on the statistics of the screw segment length distribution and scaled stress fluctuations, the cross-slip rate in CDD can then be defined following Eq. (7) as

$$\dot{r}_{cs}^{i \to j} = \left\langle \frac{\sum_{m=1}^{N} I_{cs,m}\, l_m}{\sum_{m=1}^{N} l_m} \right\rangle \frac{1}{\Delta t}; \quad I_{cs,m} = \begin{cases} 1 \text{ if } P(l_m) = cl_m e^{-\frac{E_{act}}{kT}} > R, \\ 0 \end{cases} \qquad (11)$$

where $l_m$ corresponds to the length of the $m^{\text{th}}$ sampled screw segment, $\Delta t$ corresponds to the timestep, $N$ corresponds to the number of screw segments in the unit volume, $I_{cs,m}$ is the cross-slip indicator function for the $m^{\text{th}}$ screw segment, and $R$ is a random number. The cross-slip indicator function is 1 for the $m^{\text{th}}$ segment if the cross-slip probability based on Eq. (6) is greater than the generated random number else it is zero. The functional expression for the normalization constant and activation energy varies based on the model chosen to characterize the cross-slip process. The activation energy expression ($E_{act}$) for the three different cross-slip models: Kubin (Devincre et al., 2011; Kubin et al., 1992), Hussein (Hussein et al., 2015) and Malka (Malka-Markovitz et al., 2021) are given below.

$$E_{\text{act}} = -V(|\tau_s^{cs}| - \tau_{\text{III}}) \text{ (Kubin)}, \tag{12}$$

$$E_{\text{act}} = E_{\text{cons}} - V(\tau_e^p - \tau_e^{cs}) \text{ (Hussein)}, \tag{13}$$

$$E_{\text{act}} = E_{\text{cons}}\left[\frac{G(\tau_e^p)}{2} + \frac{G(\tau_e^{cs})}{2}\left[\tanh(l_c) - \frac{2\zeta_{\text{Ls}}l_c E^*}{1.55} - \frac{\lambda}{1.55 G(\tau_e^{cs})}(\zeta_{\text{Ls}}l_c)^3\right]\right] \text{ (Malka)}. \tag{14}$$

The definitions of the parameters in the above three model equations are given in Table 1 below.

Table 1. Definitions of symbols in Eq. (12)-(14).

| Symbols | Definitions | Symbols | Definitions |
|---|---|---|---|
| $E_{\text{cons}}$ | Constriction energy | $l_c$ | Length of dislocation dissociated into CS plane |
| $V$ | Activation volume | $\zeta_{\text{Ls}}$ | Fitting parameter |
| $G(\tau)$ | $\left(1 + \frac{\sqrt{3}b}{6\gamma}\tau\right)^{-1}$ | $E^*$ | Offset of interaction energies in glide and cross-slip plane. $\left(\ln\left(\frac{G(\tau_e^{cs})}{G(\tau_e^p)}\right)\right)$ |
| $\gamma$ | Stacking fault energy | $\lambda$ | $\frac{1}{6}\left(\frac{b\tau_s^{cs}}{\gamma}\right)^2$ |

The Schmid stress and Escaig stress used in these activation energy formulations are the effective stress values on the screw segments after considering the fluctuation stress statistics. For example, in the model by Kubin, the activation energy depends on Schmid stress in the cross-slip plane ($\tau_s^{cs}$), whereas in the model by Hussein the activation energy depends on the difference between the Escaig stress on the primary plane ($\tau_e^p$) and cross-slip plane ($\tau_e^{cs}$). The effective stress for these two models after accounting for stress fluctuation can then be written in the form

$$\tau_s^{cs} = \bar{\tau}_s^{cs} * \hat{\tau}_s^{cs} + \bar{\tau}_s^{cs}, \tag{15}$$

$$\tau_e^{cs} = \bar{\tau}_e^{cs} * \hat{\tau}_e^{cs} + \bar{\tau}_e^{cs}, \tag{16}$$

where $\bar{\tau}_s^{cs}$ and $\bar{\tau}_e^{cs}$ correspond to the mean Schmid stress on the cross-slip plane and mean Escaig stress on the cross-slip plane in CDD respectively. $\hat{\tau}_s^{cs}$ and $\hat{\tau}_e^{cs}$ correspond to the scaled Schmid and Escaig stress fluctuation variables which can be sampled from the estimated PDF based on Eq. (6).

## 4. Results

In this section, the numerical experiment used to validate the newly proposed cross-slip framework, referred to as CDD cross-slip framework hereafter, is discussed first. This involves comparing the average cross-slip rates obtained using the CDD cross-slip framework and DDD methodology for a given DDD configuration. Following that, the effect of different cross-slip models on the mechanical behavior and dislocation microstructure is studied by performing bulk CDD simulations using different cross-slip models of interest.

### 4.1. Cross-slip rate comparison

The main objective behind developing a new framework for cross-slip modelling in CDD is to capture the cross-slip rates accurately in CDD. Hence, to validate the CDD cross-slip framework a discrete configuration is sampled from DDD simulation, and the cross-slip rate is evaluated using the CDD framework and compared with DDD cross-slip rate. The DDD simulation tool used to conduct this experiment is the DDD simulation code microMegas (Devincre et al., 2011). Although, the DDD simulation code uses a cross-slip model that was derived from (Kubin et al., 1992), the DDD cross-slip rate was also evaluated for Hussein and Malka's model for the chosen DDD configuration based on the data regarding screw segment length and stress on the dislocations, which are readily available.

The first step in the cross-slip rate comparison process is converting the chosen DDD configuration into a CDD configuration by performing a coarse graining process. The coarse graining process involves partitioning the domain into small volumes and smearing out the discrete dislocation within each volume to obtain a continuum representation of the discrete configuration. The details regarding the coarse graining procedure can be found in (Vivekanandan et al., 2022). Following the coarse graining procedure, the coarse-grained state variables like dislocation density and coarse grained stress can be estimated for the given DDD configuration which can then be used in the CDD framework to estimate cross-slip rate. The details regarding the slip system nomenclature used in microMegas is mentioned in Table 2.

Table 2. Slip system enumeration for FCC crystals in microMegas

| Slip system | 1 | 2 | 3 | 4 | 5 | 6 | 7 | 8 | 9 | 10 | 11 | 12 |
|---|---|---|---|---|---|---|---|---|---|---|---|---|
| Plane | (111) | (1$\bar{1}$1) | (1$\bar{1}$1) | (11$\bar{1}$) | (11$\bar{1}$) | (111) | ($\bar{1}$11) | (1$\bar{1}$1) | (111) | ($\bar{1}$11) | (11$\bar{1}$) | ($\bar{1}$11) |
| Direction | [$\bar{1}$01] | [$\bar{1}$01] | [011] | [011] | [$\bar{1}$10] | [$\bar{1}$10] | [110] | [110] | [01$\bar{1}$] | [01$\bar{1}$] | [101] | [101] |

The average DDD cross-slip rate is estimated based on the Eq. (7) whereas the average CDD cross-slip rate based on the new framework is estimated by Eq. (11). In the case of the former, the length of the discrete screw segment and stress on dislocations from DDD data are used while estimating the cross-slip probability whereas in the case of the latter, the screw segment length is sampled from the exponential distribution based on the coarse grained dislocation density and the effective stress on the dislocations is estimated based on Eq. (15) by sampling values from the Lorentzian distribution for scaled Schmid stress fluctuations and coarse grained stress. The material parameters used in these simulations are given in Table 3.

Table 3. Material parameters used in the simulations.

| **Material parameters** | **Values** | **Material parameters** | **Values** |
|---|---|---|---|
| Shear modulus ($\mu$) | 75 GPa | Stacking fault energy ($\gamma$) | 0.03 J/m$^2$ |
| Poisson's ratio ($\nu$) | 0.26 | Fitting parameter ($\zeta_{Ls}$) | 0.6 |
| Drag coefficient ($B$) | $25 \times 10^{-5}$ Pa s | Constriction energy ($E_{cons}$) | 1.55 eV |
| Burgers vector ($b$) | 0.254 nm | Activation volume ($V$) | 1800b$^3$ m$^3$ |
| Stress at beginning of stage III hardening ($\tau_{III}$) | 52 MPa | Temperature ($T$) | 300 K |

Fig. 4 shows the average cross-slip rate computed for a DDD configuration at 0.5% strain using the CDD cross-slip framework and DDD method for the three cross-slip models. The cross-slip rates for the case of CDD framework are evaluated for 1000 different realizations which were sampled from the screw segment length distribution and the stress fluctuation distribution. The mean of those 1000 different instances is denoted by the black '◊' marker along with a bar that indicates the maximum and minimum values in Fig. 4. The cross-slip rate obtained using DDD method is represented by the red '○' circle. It is evident from the figure that the cross-slip rates predicted by the CDD framework can take a range of values fixed by the sampled statistics. In addition to that, it can also be observed that the spectrum of cross-slip values spans a wider range for the case of inactive slip systems, denoted by the blue color in Fig 4, compared to the active ones. The reason for this is that the dependence of activation energy on sampled stress fluctuation statistics is more prominent due to the lack of resolved shear stress along the inactive slip systems. Consequently, any changes in the stress fluctuations affect the cross-slip rate more significantly in the case of inactive slip systems compared to active slip systems. Furthermore, comparing the CDD cross-slip rates with DDD cross-slip rates for all three models, it can be observed that in most of the cases the DDD values lie within the spectrum of CDD values. This indicates that the CDD model is indeed able to capture the cross-slip behavior in DDD based on the mean field variables and DDD statistics. A better agreement between the two methods can be expected when an ensemble of DDD configuration at the same strain level is sampled from multiple DDD simulations so that different possible local states are explored, which can then be averaged to compare the results from CDD.

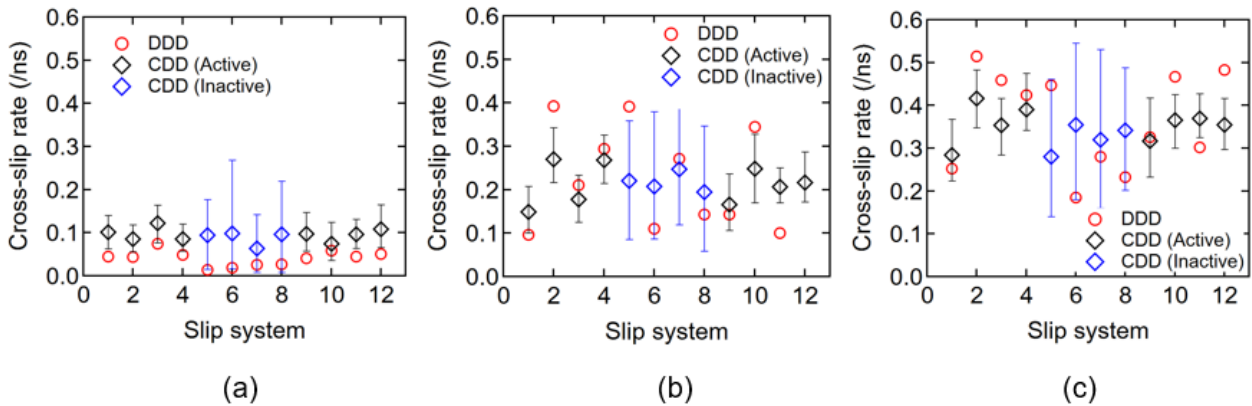

Fig. 4. Average cross-slip rate computed using (a) Kubin, (b) Hussein and (c) Malka's model based on DDD and CDD framework for a DDD configuration sampled at 0.5% strain.

Comparing the cross-slip rates given in Fig. 4 among the three different models shows that Malka's model yields the highest cross-slip rate whereas the Kubin's model has the lowest cross-slip rate. The difference in the cross-slip rates between the three models can be attributed to the stress components considered in the activation energy expression as well as the parameters used to calibrate the model. For instance, the Kubin's model considers only the Schmid stress along the cross-slip plane to determine the cross-slip activation energy barrier, which was shown to have the least effect on reducing the cross-slip activation energy compared to other stress components (Kang et al., 2014). Consequently, one can expect it to have the lowest cross-slip rates amongst other models. Additionally, since Kubin's model was calibrated based on the cross-slip activity at the beginning of stage III hardening, high stress levels are required to initiate cross-slip compared to other models which does not have such limitations. Similarly, Hussein's model considers the cross-slip activation energy to be linearly dependent on the difference between the Escaig stresses on the primary and cross-slip plane. According to this model, cross-slip is never possible when Escaig stress on the cross-slip plane is higher compared to that of the primary plane. However, the line tension model by (Malka-Markovitz and Mordehai, 2019) showed that when the effect of Schmid stress on the cross-slip plane is considered, cross-slip is possible even though the Escaig stress on the cross-slip plane is higher compared to that of the primary plane. Since Malka's model considers the effect of all stress components the chances for a dislocation to cross-slip is higher in their model compared to the other two models, which is evident from Fig. 4.

4.2. Effect of different cross-slip models on bulk CDD simulations

The aim of this section is to study the effect of different cross-slip activation energy formulations on the mechanical response of the dislocation system and the dislocation microstructure evolution. Hence, CDD simulations using Kubin's model (Kubin et al., 1992), Hussein's model (Hussein et al., 2015) and Malka's model (Malka-Markovitz et al., 2021) were performed based on the new framework for the monotonic loading case. All simulations were performed using the material properties listed in Table 3.

The simulation domain size in these simulations is 5µm × 5µm × 5.303µm and the applied strain rate is $20s^{-1}$ along [001] direction. The slip plane normal and Burgers vectors of all the slip systems for a FCC crystal in CDD is given in Table 4. The initial dislocation configuration was made up of 10 circular dislocation loops with radius ranging from 2µm and 5µm in each slip system. The initial total dislocation density used in these simulations is $1.5 \times 10^{12}$ m$^{-2}$.

Table 4. Slip system enumeration for FCC crystals in CDD

| Slip system | 1 | 2 | 3 | 4 | 5 | 6 | 7 | 8 | 9 | 10 | 11 | 12 |
| --- | --- | --- | --- | --- | --- | --- | --- | --- | --- | --- | --- | --- |
| Plane | (111) | ($\bar{1}$11) | ($\bar{1}$11) | (11$\bar{1}$) | (11$\bar{1}$) | (1$\bar{1}$1) | (1$\bar{1}$1) | (111) | (111) | (11$\bar{1}$) | ($\bar{1}$11) | (1$\bar{1}$1) |
| Direction | [0$\bar{1}$1] | [0$\bar{1}$1] | [101] | [101] | [011] | [011] | [$\bar{1}$01] | [$\bar{1}$01] | [$\bar{1}$10] | [$\bar{1}$10] | [110] | [110] |

Fig. 5(a) shows the average cross-slip rate, $\dot{r}_{cs}^{1 \rightarrow 2}$, defined in Eq. (11) for slip system 1 for the three models obtained by averaging the cross-slip rate throughout the CDD simulation domain. The average cross-slip rate obtained from the Kubin's model is significantly less compared to that of Hussein and Malka's models. This can be attributed to the origins of Kubin's model, which was discussed in the previous section. Fig. 5(b) shows the average cross-slip rate $\dot{r}_{cs}^{2 \rightarrow 1}$ for slip system 2, which is the collinear slip system of 1, for all three models. Comparing Figs. 5(a) and 5(b), it can be observed that the average cross-slip rates of collinear slip systems are similar for both Kubin and Malka's model, but the behavior is markedly different for the case of Hussein's model. The difference in behavior for the case of Hussein's model stems from the bias which the activation energy expression enforces on the cross-slip probability. Since the activation energy depends on the difference between Escaig stress in both the glide and cross-slip plane, as can be seen from Eq. (14), it is expected that if cross-slip is favorable from slip system 1 to slip system 2 at a point, then cross-slip from slip system 2 to 1 would be less favorable.

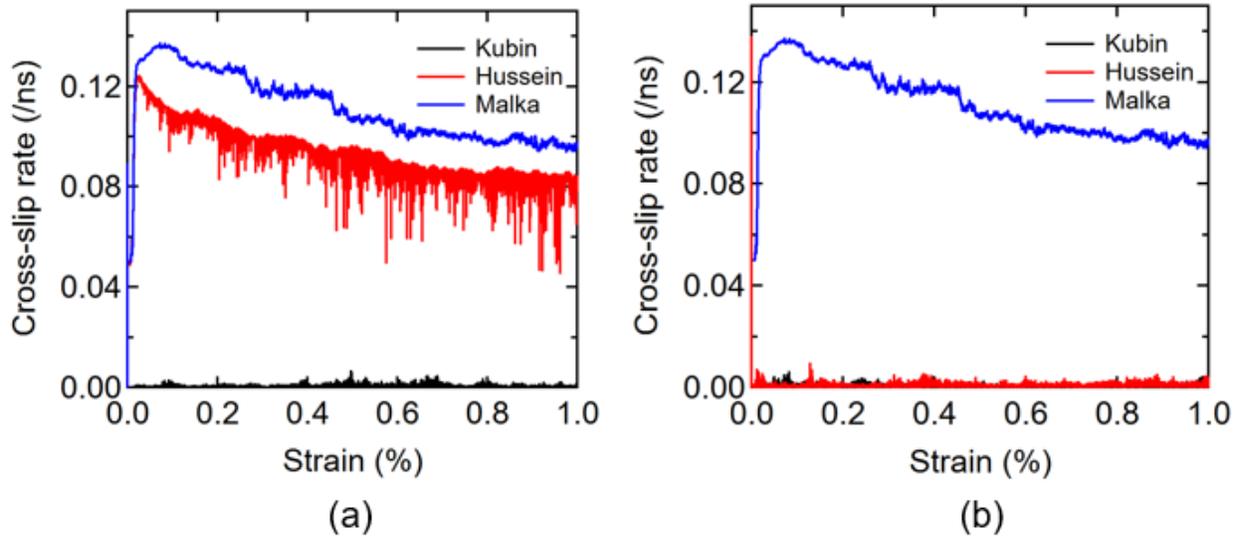

Fig. 5. Average cross-slip rates of (a) slip system 1 and (b) slip system 2 for three different models in CDD simulation.

Figs. 6(a) and (b) show the stress-strain curves and dislocation density evolution curves of all the three models, respectively. The stress-strain response in Fig. 6(a) shows that there is a difference in the stress at which yielding occurs and the rate of hardening between different models, although it is minimal. Comparing Fig. 5 and 6 shows that the models with higher cross-slip rates have lower yield stress and higher rate of hardening. This can be explained by observing Fig. 6(b) which shows that for models with higher cross-slip rate dislocation density initially evolves slower and then starts to grow faster. Consequently, we observe for models with higher cross-slip rate, the yield stress is lower due to fewer dislocations and the rate of hardening is higher due to faster increase in dislocation density.

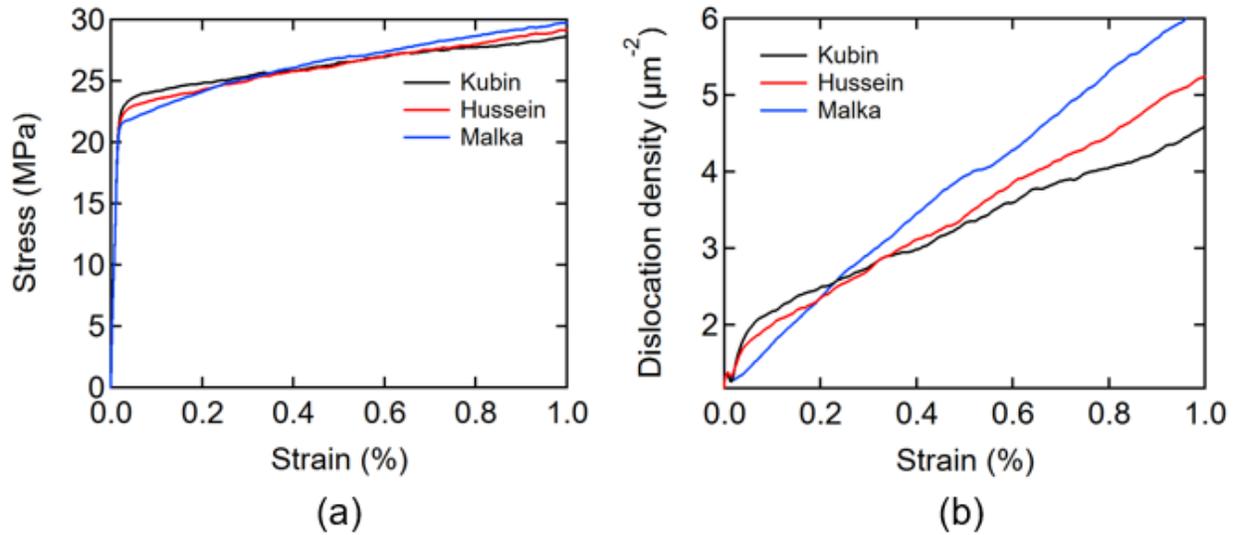

Fig. 6. (a) Stress-strain curves and (b) total dislocation density curves obtained from CDD simulation for the three different models.

Fig. 7 shows the slip system dislocation density evolution for all three models. As expected, the dislocation density for an active slip system (SS 1-8) evolves faster compared to the inactive slip system (SS 9-12) in all cases. But in the case of Hussein's model, a qualitative difference in the slip system density behavior is observed compared to other two models. The active slip systems evolve as two separate groups as can be seen in Fig. 7(b) as opposed to one in the other cases. This is due to the discrepancy in the cross-slip rates, which was shown in Fig. 5, because of which active slip systems with higher cross-slip rates evolved slower compared to their collinear counterparts since more dislocations were transferred to their collinear counterpart from them rather than to them. It is important to note that the same effect is not observed in the case of inactive slip systems since the dislocation density increase in these cases mainly comes from the glissile junction interaction between the active slip systems.

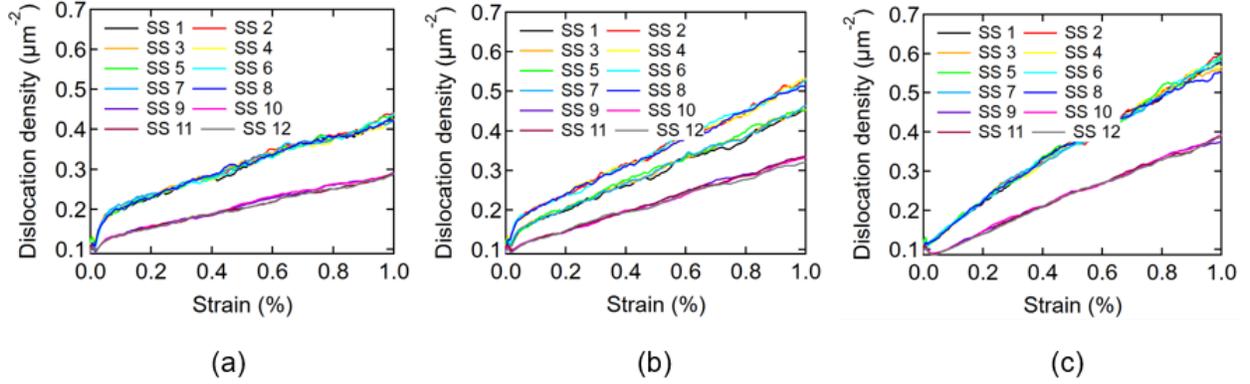

Fig. 7. Slip system dislocation density curves of simulations (a) based on Kubin's model (b) based on Hussein's model, (c) based on Malka's model.

### 4.3. A dislocation relaxation experiment

A relaxation experiment was performed wherein the dislocation configuration is loaded to a strain level of 0.25% in a single time step, at a strain rate of the order $10^6$ s$^{-1}$ along [001] direction and then allowed to relax. Loading the simulation domain to 0.25% strain in a single time step results in high stress levels since the applied strain gets converted mostly to elastic strain as the dislocations do not have enough time to move. Once the high stress is induced in the domain, the system is then allowed to relax by maintaining the mean strain at 0.25% subjected to periodic boundary conditions. This approach was chosen to study the dislocation microstructure evolution at high stress levels because accessing such stress values through typical monotonic loading simulations is near impossible due to the computational constraints and strain levels required to reach such stresses.

The relaxation experiment described above was conducted for the three cross-slip models along [001] orientation. The simulation box size used in this simulation was $2.5\mu m \times 2.5\mu m \times 2.75\mu m$ and the initial dislocation density was chosen to be $3 \times 10^{14}$ $\mu m^{-2}$. The initial dislocation configuration was made up of 10 circular dislocation loops with radii ranging from $2\mu m$ and $3\mu m$ in each slip system. The material parameters used in these simulations are listed in Table 3.

Fig. 8 shows the stress-time curve and dislocation density behavior over the course of the relaxation experiment. The initial steep increase in stress in Fig. 8(a) corresponds to the applied load stored as elastic strain in the simulation domain. With the progression of time, the stress in the simulation domain reduces due to the relaxation and then reaches a stable state eventually for all three models. Fig. 8(b) shows the dislocation density evolution with time for the three models. The dislocation density decreases asymptotically due to the dislocation annihilation that happens during the relaxation tending towards a stable state for all three models.

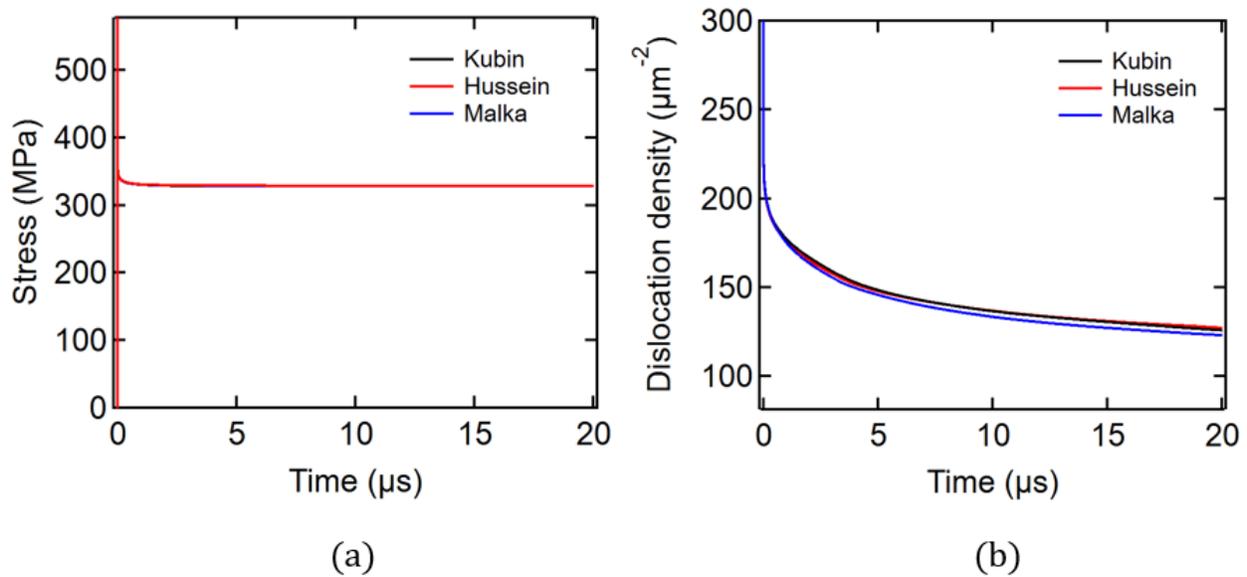

Fig. 8. (a) Stress-strain curve and (b) total dislocation density evolution curve of the relaxation experiment.

Fig. 9 and 10 shows the dislocation microstructure obtained from the relaxation experiment for all the three models by visualizing the scalar dislocation density after applying the threshold filter over the whole simulation domain and along (010) plane respectively. The threshold filter assigns a color, in this case black or white, based on the dislocation density value at a given point. For the Figs. 9 and 10, if the dislocation density value at a point is greater than 175 $\mu m^{-2}$, then the color black was assigned else the color white was assigned. After applying the threshold filter, the cell-type structures are observed along the planes (110) and (010), as observed in experiments (Ungar et al., 1984).

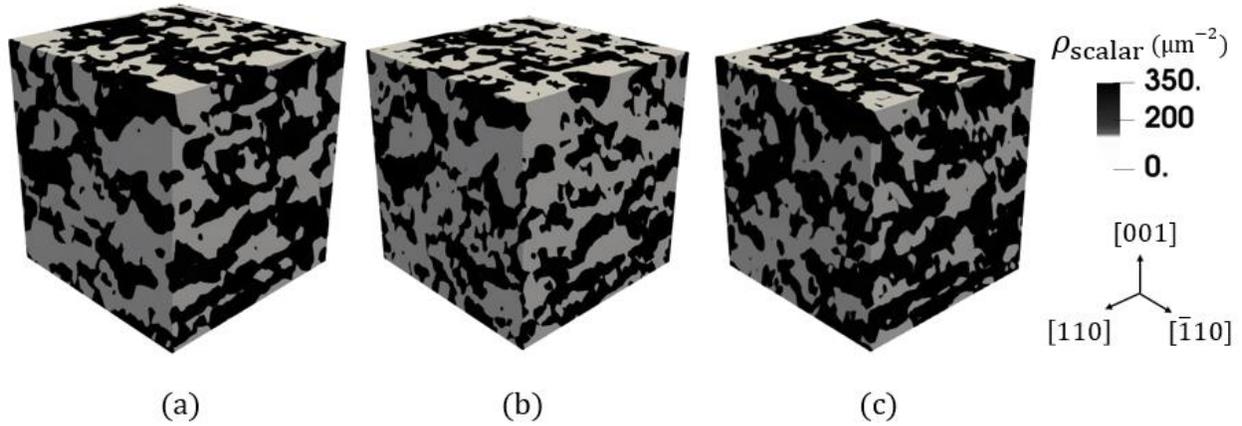

Fig. 9. Dislocation microstructure obtained from the CDD relaxation experiment by visualizing the scalar dislocation density after applying the threshold filter using (a) Kubin's model, (b) Hussein's model and (c) Malka's model for modeling cross-slip process. The threshold filter has been applied to accentuate the features present in the image.

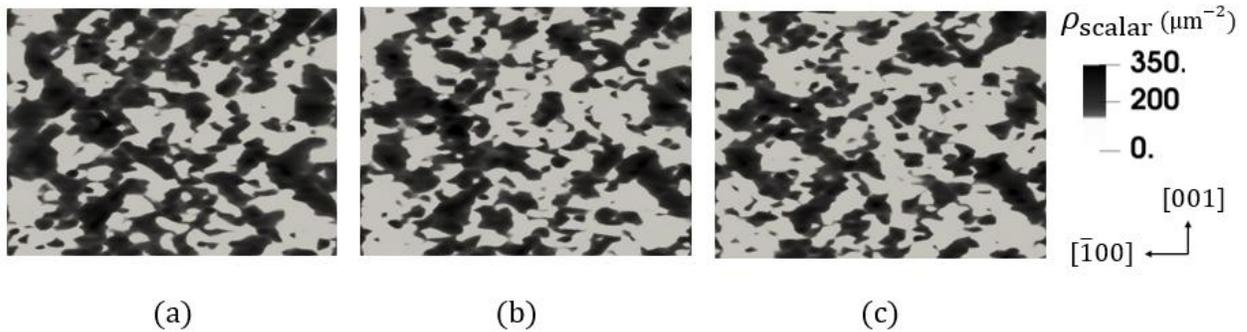

Fig. 10. Dislocation microstructure from the CDD relaxation experiment along (010) plane obtained by visualizing the scalar dislocation density after applying the threshold filter using (a) Kubin's model, (b) Hussein's model and (c) Malka's model for modelling cross-slip process. The threshold filter has been applied to accentuate the features present in the image.

To quantify the similarities and differences between dislocation microstructures predicted by CDD for the three cross-slip models of interest, autocorrelations of the 2D dislocation microstructure images in Fig. 10 were computed and compared. These autocorrelations were computed for single cross-sections of the simulation volume from the filtered data using FFT algorithm (Frigo, 1999). Fig. 11 (a), (b) and (c) shows the autocorrelation for the corresponding three images in Fig. 10. The autocorrelations of these images characterize the probability of finding dislocations at a prescribed distance and direction from a randomly placed point on the image. The X and Y axis in the images represent the distance in nm and the color bar captures the probability. All the three images show a bright red spot with the highest probability at the center and the probability decreases uniformly along all directions approximately which is evident from the gradual change in the color as one goes away from the center. Fig. 11(d) shows the radial pair correlation function for all three dislocation microstructure images in Fig. 10 obtained using MATLAB (Gavagnin et al., 2018), which decreases with increase in radial distance till it reaches a minimum and then rises again before becoming a constant for all three models. It is evident from the figures that the dislocation microstructure morphology of all three models is very similar as all the models follow the same trend.

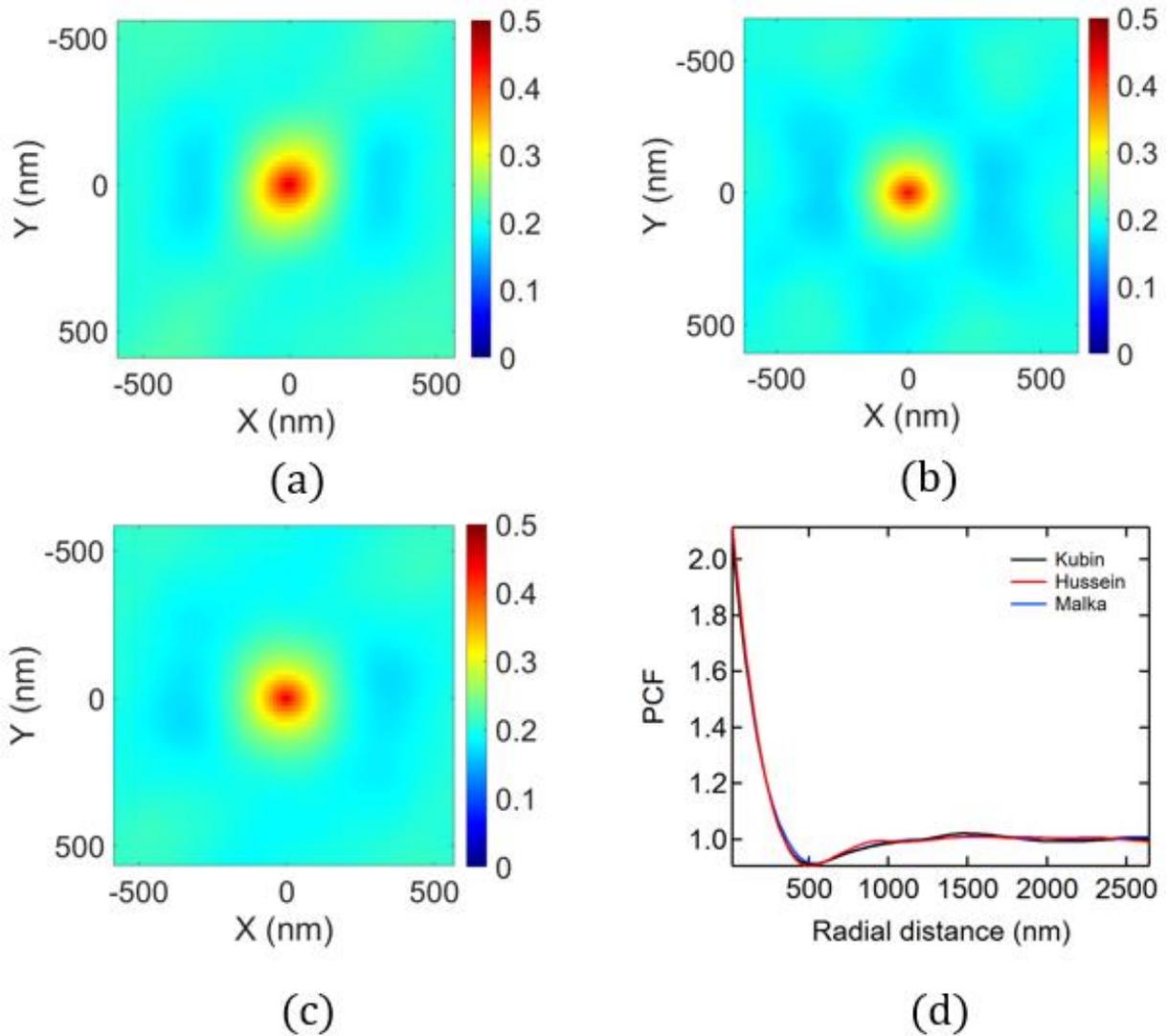

Fig. 11 Spatial autocorrelation of dislocation microstructure in Fig. 10 obtained using FFT algorithm (Frigo, 1999) for (a) Kubin's model, (b) Hussein's model and (c) Malka's model. (d) Radial pair correlation function for all three models.

## 5. Discussion

The results presented in the previous section highlight the capability of the new CDD cross-slip framework which represents an improvement over the other existing approaches used to model cross-slip in CDD. The cross slip modeling approach proposed here incorporates the information about local coarse grained stress, dislocation density, the screw segment length statistics, and stress fluctuation statistics, thus allowing to overcome the shortcomings of other approaches used to capture the cross-slip process in CDD (Deng and El-Azab, 2010; Sudmanns et al., 2019; Xia et al., 2016). For instance, in the works of (Deng and El-Azab, 2010; Xia et al., 2016), the cross-slip rate was estimated by sampling the averaged cross-slip rate time series data from DDD simulation after performing time coarse graining analysis. Although, this approach made use of the cross-slip statistics from DDD, it was averaged over the simulation domain thereby ignoring the effects of the local stress and its fluctuations. This issue was addressed in the proposed model by making use of the local stress and dislocation density information, which was used to sample DDD statistics like screw segment length and stress fluctuation statistics to estimate the cross-slip rate. Likewise, the work of (Sudmanns et al., 2019) estimated the cross-slip rate by using the local coarse grained stress directly in the DDD cross-slip probability expression (Kubin et al., 1992), without incorporating the effects of local dislocation arrangement which was shown to result in the markedly different local stress states (Vivekanandan et al., 2022). The new CDD cross-slip framework addressed this issue by making use of the screw segment length statistics and the stress fluctuation statistics which captures the different possible underlying local state for a given coarse grained description.

The validity of the newly proposed cross-slip modeling approach was tested by comparing the cross-slip rate estimated using CDD and DDD methodology for a chosen DDD configuration. Cross-slip rate estimated using DDD methodology made use of the stress on the screw dislocations and screw segment length whereas the cross-slip rate obtained by using the CDD formalism used the coarse-grained stress and dislocation density in addition to the values sampled from DDD statistics like the screw segment length and stress fluctuation distribution. The sampled values correspond to one of the many possible configurations of the ensemble and hence it is indeed expected that the estimated cross-slip rate will vary along a spectrum and will not be an accurate match to the actual cross-slip rate. Nevertheless, Fig. 4 showed that CDD cross-slip rate estimated using the new framework is within reasonable limits of DDD cross-slip rate for all three models. For an accurate comparison between the two methods, a set of DDD configurations should be sampled from multiple DDD simulations such that different local configurations for a given strain level can be accounted for during cross-slip rate estimation. The average cross-slip rate of the ensemble can then be used to compare with the average of CDD cross-slip rates that were obtained by sampling statistics from the screw segment length and stress fluctuation distribution.

The effect of different cross-slip activation energy formulations on the mechanical behavior and dislocation microstructure was investigated in this work. It was shown that the different cross-slip activation energy formulations indeed resulted in different average cross-slip rates, but the effect of the different cross-slip rates on the stress-strain response and dislocation density evolution was shown to be minimal in Fig. 6 and 7. A plausible explanation for this behavior is the nature of the loading orientation, which is [001] loading orientation and type of loading used in the simulations, which is monotonic loading and cross-slip rates of collinear slip systems. The [001] loading orientation activates 8 slip systems comprising 4 pairs of collinear slip systems sharing the same Burgers vector for FCC crystal such that they all have the same Schmid factor. Similarly, the remaining 4 inactive slip systems comprising of 2 pairs of inactive slip systems also have the same Schmid factor. Hence, for the case where the loading is monotonic and collinear slip systems having similar cross-slip rates, one can expect dislocations evolving in a similar manner despite different cross-slip rates due to the dominance of the applied component of the Schmid stress. Consequently, the difference in the stress-strain response can also be expected to follow similar trend for models with different cross-slip rates. The relaxation experiment which was devised to study how dislocations reorganize themselves at higher stress levels and dislocation density also yielded similar stress-strain and dislocation density evolution behavior for all three models. With regard to the collective behavior of dislocations, the difference between the three models in this case was almost non-existent compared to the monotonic case indicating that at higher stresses, the differences between the models become insignificant. Finally, despite subjecting the dislocation system to a high strain rate loading before relaxation, the dislocation microstructure showed cell-like type structure typically observed in monotonic simulation for all three models.

## 6. Conclusion

A data driven approach to model cross-slip behavior in CDD was proposed. The model made use of the screw segment length distribution and stress fluctuation statistics from DDD to inform CDD model about the essential information required for cross-slip modelling which was lost due to coarse graining. It was shown that the screw segment distribution in DDD follows the exponential distribution and the stress fluctuation statistics for screw segments follows the Lorentzian distribution. Using these statistics from DDD, it was shown that the average cross-slip rate over the whole domain obtained using the new framework was able to match the DDD cross-slip rate. In addition, the effect of different activation energy formulations on the stress-strain response, dislocation density evolution and dislocation microstructure were studied. It was found that, although the average cross-slip rates of the different cross-slip activation energies were different, the effect on stress-strain response and dislocation density evolution were minimal. Furthermore, it was shown that in the presence of high local stress and dislocation density, the new CDD framework was able to capture the dislocation self-organization in cell-like structure typically observed along the {100} and {110} type planes in experiments.

## 7. Acknowledgments

The authors are grateful for the support from the Naval Nuclear Laboratory, operated by Fluor Marine Propulsion, LLC for the US Naval Reactors Program. The computational methodology and writing of the work were partly supported by the National Science Foundation, Division of Civil, Mechanical, and Manufacturing Innovation (CMMI), through award number 1663311 and by the US Department of Energy, Office of Science, Division of Materials Sciences and Engineering, through award number DE-SC0017718 at Purdue University.